\documentclass[%
 aip,
 jmp,%
 amsmath,amssymb,
 reprint,%
]{revtex4-1}

\usepackage{graphics}
\usepackage{graphicx}
\usepackage{dcolumn}
\usepackage{bm}

\begin{document}

\preprint{AIP/123-QED}

\title{Anomalous viscosity of vortex hall states in graphene}
\thanks{Corresponding author. Email: mrabiu@uds.edu.gh.}

\author{M. Rabiu}
 \email{mrabiu@uds.edu.gh}
 \affiliation{Department of Applied Physics, Faculty of Applied Sciences, University for Development Studies, Navrongo Campus, Ghana.
}

\author{S. Y. Mensah}%
\affiliation{Department of Physics, Laser and Fiber Optics Center, University of Cape Coast, Cape Coast, Ghana.
}%

\author{S. Y. Ibrahim}
 \affiliation{University for Development Studies, Faculty of Mathematical Science, Department of Mathematics, Navrongo, Ghana.
}

\author{S. S. Abukari}
\affiliation{Department of Physics, Laser and Fiber Optics Center, University of Cape Coast, Cape Coast, Ghana.
}%

\date{\today}

\begin{abstract}
We study temperature effect on anomalous viscosity of Graphene Hall fluid within quantum many-vortex hydrodynamics. The commonly observed filling fractions, $\nu$ in the range $0 < \nu < 2$ is considered. An expression for anomalous viscosity dependent on a geometric parameter-Hall expansion coefficient, is obtained at finite temperatures. It arises from strained induced pseudo-magnetic field in addition to an anomalous term in vortex velocity, which is responsible for re-normalization of vortex-vortex interactions. We observed that both terms greatly modify the anomalous viscosity as well as an enhancement of weakly observed v fractions. Finite values of the expansion coefficient produce a constant and an infinite viscosities at varying temperatures. The infinities are identified as energy gaps and suggest temperatures at which new stable quantum hall filling fractions could be seen. This phenomenon is used to estimate energy gaps of already measured fractional quantum Hall states in Graphene.
\end{abstract}

\maketitle

\section{Introduction\label{Sec:Section1}}
Graphene is a monolayer of carbon atoms tightly packed into a two-dimensional (2D) honeycomb lattice. Since its discovery in 2004, graphene has attracted a great deal of attention mainly due to its exceptionally high crystal and electronic quality. Shear viscosity has been studied in graphene~\cite{Mueller2009, Mendoza2013}. The shear viscosity establishes transverse velocity gradients  that obstruct coherent electron flow. Under some conditions, one has relativistic particles in graphene forming a quantum Hall fluid; a strongly correlated sates of matter, which flows just like fluid and without such shearing resistance or dissipation. The  viscosity measured in the quantum Hall fluid is what is termed Anomalous (also called Hall or dissipationless) viscosity~\cite{Wiegmann2013}. Quantum fluids are particularly interesting especially due to the remarkable natural phenomenon occurring in superconductivity, superfluidity, ultracold atoms. Fractional Quantum Hall (FQH) Effect  is yet another example.  The ground states of FQH states are holomorphic in nature and gapped. These gapped states are characterized by a universal anomalous viscosity. The viscosity is revealed when stress tensor becomes sensitive to  stress preserving deformations of the fluid.  In this context, the origin of the anomalous term is due to fluid velocity diverging at microscopic scale which deforms a metric locally and causes dilatation of particle coordinates. The divergences due to individual particles are collectively manifested at macroscopic scale as an anomalous term. Another important term arising from metric deformations at finite temperature is Hall expansion coefficient. It is well established that  graphene can respond to local deformations by producing strain which in turn induces giant pseudo-magnetic fields as much as $10\, T$ in strained graphene~\cite{Guinea2010} and $300\, T$ in graphene nanobubbles~\cite{Levy2010}. The induced field allows vortices in the system to feel an effective magnetic field. It is the Hall expansion coefficient that captures the contribution.

Recently, there has been a great deal of interest and renewed focus on the anomalous viscosity, $\eta_A$ of quantum fluids. In particular, ref~\cite{Read2009} obtained a universal relation $\eta_A = (\hbar\rho/4)\mathcal{S}$ for FQH states of generic bulk systems by studying the response to metric deformations.  In Ref.~\cite{Siavah2014} $\eta_A$ ($= (B/8\pi)\kappa$) was obtained for the FQH  states of relativistic particles in graphene by electromagnetic and gravitational response. The same general result, including anomalous term, was also obtained by Abanov within effective hydrodynamic theory~\cite{Abanov2013} and within Euler vortex hydrodynamics~\cite{Wiegmann2013}. However, all these results lack finite temperature effects which can have remarkable consequence on $\eta_A$ and other transport properties.

In the following, we consider flows of quantized Hall vortices in the graphene Hall fluid as elementary objects forming themselves a highly correlated quantum fluid. In the regime of long-wave slow motion, an accurate hydrodynamic description becomes possible and does, in its own validity, not depend on the microscopic behavior of the electronic fluid. Vortex-vortex interactions will then be responsible for appearance of vortex FQH Effect. The hydrodynamics of vortex matter presented here differs from Euler hydrodynamics \cite{Stone1990} by an anomalous term. Within this many-vortex approach, the anomalous term has been derived for bulk single component non-degenerate systems \cite{Wiegmann2013} without the Hall expansion coefficient.

The remaining of the paper is organized as follows. In Section \ref{Sec:Section2} we obtained vorticity velocity equation from Euler hydrodynamic equation through Boltzmann transport equation for graphene. We obtained a quantized Helmholtz-Kirchoff vortex solution. Vortex flux and momentum conservation laws are subsequently derived. The stress tensor is deduced from which the anomalous viscosity is read off. In  Section \ref{Sec:Section3}, we analyzed behavior of the anomalous viscosity under density and temperature profiles for different filling fractions. We concluded in  Section \ref{Sec:Section3} highlighting possible applications of our results.

\section{\label{sec:The-Theory} Theoretical model\label{Sec:Section2}}
\subsection{Euler hydrodynamics and point vortices}
Two dimensional Euler hydrodynamics can be straightforwardly derived following Boltzmann transport equation at local equilibrium \cite{Camiola2014}. Particle distribution is reduced to continuity and Euler equations for density and momentum
\begin{equation}
 D_t\rho^{\alpha} = 0\quad \mbox{ and }\quad D_t{\bf u}^{\alpha} + \nabla {\rm p}^{\alpha}= 0,\label{Eq:EFluid}
\end{equation}
respectively. Where $D_t \equiv \partial_t + {\bf u}\cdot {\bf \nabla}$ and ${\bf u}$ is macroscopic fluid velocity connected with the microscopic electron velocity, ${\bf v} = v_F{\bf k}/k$. ${\rm p}$ is the partial pressure per density and $\alpha$ is the fluid component index. For graphene, $\alpha\equiv$ (K$^\prime\uparrow$, K$\uparrow$, K$^\prime\downarrow$, K$\downarrow$). ${\rm K, K^\prime}$ and $\uparrow,\downarrow$ are the valley and spin indexes, respectively. Taking the curl of the Euler equation (\ref{Eq:EFluid}), we get
\begin{equation}
  D_t\omega^{\alpha} = 0.\label{Eq:VFluid}
\end{equation}
Where the quantity $\omega = {\bf \nabla}\times {\bf u}$ is vorticity which is non-zero for rotational incompressible flows. $\omega$ is considered as frozen into the fluid and constitute its own fluid. The continuity equation for the vortex fluid $ D_t\rho^{\alpha}_{\rm v} = 0$, also holds. A solution for equation (\ref{Eq:VFluid}) exists and consists of point-like vortices. The fluid velocity is
\begin{equation}
  {\rm u}^{\alpha} = -i\Omega\bar{z}^{\alpha} + i\sum\frac{\Gamma^{\alpha}_i}{z^{\alpha} - z_i^{\alpha}(t)}. \label{eq:Fluidvelo}
\end{equation}
Where $z = x + {\rm i} y$, $u = u_x - {\rm i} u_y$, $\Gamma$ is the circulation and $\Omega$ which is identified with cyclotron frequency of the fluid particles. Assuming a flow in which the circulation, $\Gamma^{\alpha}_i$ ($=\Gamma^{\alpha}$) is both minimal and chiral so that in the limit, $N_{\rm v} \to \infty$ rotation is compensated by the large number of vortices. If magnetic field is present, the vortices of all components are smoothly distributed with fixed mean density, $\rho_0 = (1/\pi)(\Omega/\Gamma)$. Borhn-Summarfelt phase-space quantization leads to quantization of the circulation $m_{\text{v}}\Gamma^{\alpha} = \beta^{\alpha}\hbar$. Where $m_{\text{v}}$ is inertia vortex flow and $\beta^{\alpha}$ is an integer. The equation for the slow motion of  vortices is easily obtained from Eq. \ref{eq:Fluidvelo}. It is given by 
\begin{equation}
  \text{ v}_i^{\alpha}(t) = -i\Omega\bar{z}_i^{\alpha}(t) + i\sum_{i\neq j}\frac{\Gamma^{\alpha}}{z_i^{\alpha}(t) - z_j^{\alpha}(t)}. \label{Eq:Vortex}
\end{equation}
Equation (\ref{Eq:Vortex}) is the Helmholtz-Kirchoff equation for vortices. We will specialize in the zero energy state of the system where K and K$^\prime$ components of the fluid decouple. Dynamics are then localized in either sublattice and the $\alpha$ dependence may be dropped.

\subsection{Temperature effect on dynamics of hall vortices and viscosity anomaly} 
In the quantum Hall regime of dissipationless flow, fluid particles do not carry heat flux, but vortices do. Vortices move in response to temperature gradient, $\nabla T$. This can be captured in a vortex density defined through momentum flux as ${\rm P}(r,T) = m_{{\rm v}}\{\rho_{\rm v}(r,T),{\rm v}\}/2$
\begin{equation}
  {\rm P}(r,T) = \frac{m_{\rm v}}{2}\Big\{\rho_{\rm v}(r,T),{\rm v}\Big\}.\label{eq:MomFlux1}
\end{equation}
Using the identity $\pi\delta(r) \equiv \bar{\partial }(1/z)$ and the ward identity~\cite{Wiegmann2013}  $\sum_{i\neq j}[2/(z-z_i)(z_i-z_j)] \equiv [\sum(1/(z-z_i)]^2 - \sum[1/(z-z_i)]^2$ in Eq.~\ref{eq:MomFlux1}, we get
\begin{equation}
  {\rm P}(r,T) = m_{\rm v}\rho_{\rm v}(r,T) \left[{\rm u}+\frac{\Gamma}{4}\nabla^*{\rm log}\,\rho_{\rm v}(r,T)\right]. \label{Eq:Flux}
\end{equation}

Equation (\ref{Eq:Flux}) is crucial in this studies. In particular, our results is based on the second term which dictates discussions that follows. The term is responsible for anomalous behavior of the fluid when approaching a vortex. It is a quantum or micro-scale phenomenon which manifest itself at classical regime due to possible broken translation symmetry associated with lattice scale deformations. Its presents renormalizes the vortex-vortex interactions in equation (\ref{Eq:Vortex}). It also creates stresses perpendicular to the fluid flow with no work nor dissipation. The associated transport coefficient (viscosity) is expected to be dissipationless. The momentum conservation following from broken translation invariance in the presence of external forces yields 
\begin{equation}
  \dot{{\rm P}}_a + \nabla_b\Pi_{ab} = \rho_{\rm v}F_a.
\end{equation}
Where the stress tensor is $\Pi = m_{\rm v}\rho_{\rm v}{\bf u}\cdot {\bf u} +\eta_A(\nabla^*{\bf u} + \nabla{\bf u}^*)$ $\nabla^* (\cdot)\equiv  \nabla\times (\cdot)$. $F =E + {\rm v}\times B$ is the external force. The kinetic coefficient $\eta_A$ is the anomalous (dissipationless) viscosity. Except for the temperature dependence, the dessipationless viscosity has the same structure as the universal relation obtained in literature for graphene Fractional quantum Hall states \cite{Siavah2014}. i.e,
\begin{equation}
  \eta_A(r,T) = \frac{m_{\rm v}\Gamma}{4}\rho_{\rm v}(r, T).\label{Eq:AnomalousV}
\end{equation}

In order to have a complete description of our system, one needs to quantize Eq.~\ref{Eq:Flux}. It leads to physical interpretation of the quantized circulation as the filling factor of the fractional vortex Hall states \cite{Wiegmann2013}. A connection between $\Gamma ( =(\hbar/m_{\rm v})\mathcal{S}$) and the so called shift, $\mathcal{S}$ of a Hall state is easily established. For Graphene, $\mathcal{S} =\sum_{\alpha}(\beta^{\alpha}-1)/4$. Except the re-definition of $\Gamma$, the quantized vortex velocity assumes the same form as Eq.~\ref{Eq:Flux}.
Taking the curl and and utilizing the relation $\nabla\times {\rm u} = 2\pi\Gamma(\rho_{\rm v} - \rho_0 )$, we have

\begin{equation}   
\nabla\times {\rm v} = 2\pi\Gamma\Big[\rho_{\rm v}(r,T) - \rho_0  + \frac{\eta_A(r,T)}{4\pi\Gamma}\Delta^*{\rm log}\, \rho_{\rm v}(r,T)\Big]. \label{Eq:QuantumCurVel}
\end{equation}

\subsection{Effective magnetic field}
As we have pointed out in the introduction, geometric deformations can lead to induction of giant synthetic magnetic fields. Vortices feel this contribution in addition to external magnetic field as an effective field, $B_{\rm v}$. The anomalous viscosity for the graphene-vortex system takes the shape, $\eta_A(r, t,T) = (e\kappa/8\pi)B_{{\rm v}}(r,t,T) $ with $ \kappa= \nu_G\mathcal{S}$ and $\rho = \nu_G(eB/h)$. The Fourier transform of $B_{{\rm v}}$ in time-domain yields frequency-dependent viscosity
\begin{equation}
  \eta_A(r,\omega,T) = \frac{e\kappa}{8\pi}B_{\rm v}(r,\omega,T).\label{eq:Eta}
\end{equation}
Where $r= |z_i-z_j|$. To compute $B_{\rm v}(r,\omega,T)$, we decompose it to $B_{\rm v}(\omega)B_{\rm v}(r, T)/B_0$. $B_0$ is the applied field. 

First we need to compute $B_{\rm v}$ from Eq.\ref{Eq:QuantumCurVel}. To do that, we require that in the ground sate $\nabla\times{\rm v}=0$ which yields $B_{\rm v}(r,T) = B_0 -(\hbar/4e)((1/\nu_G) - 2)\Delta^*{\rm log}\,B_{\rm v}(r,T) $. Expanding $B_{\rm v}$ in small temperature gradients to get an iterative expression for the magnetic field,
\begin{equation}
B_{\rm v}(r,T) = B_0 - \frac{\xi\hbar}{2e}\left\{\Delta^*{\rm log} B_{\rm v}(r) + \Delta^*{\rm log}\Big[1- \gamma \nabla T(r)\Big]\right\}.\label{Eq:VortedField}
\end{equation}
Where $\xi = 1/2\nu_G - 1$ is the anomalous term. The Hall expansion coefficient is defined as $\gamma = -(1/B_{\rm v})(\partial B_{\rm v}/\partial T)$. The minus sign is inferred from recent experimental results on graphene at low-temperatures~\cite{Vibhor2010, Duhee2011, Linas2014}. Because strain in graphene induces giant synthetic magnetic fields, at finite temperatures $\gamma$ can have great consequences on the system's electronic transport properties. Thus, the coefficient can characterize stress and knowing it can be very critical in strain engineering. 

Finally, to obtain $B_{\rm v}(\omega)$ we recall that the Magnus force acting on vortices has th form
\begin{equation}
  \rho_{\rm v} F=m_{\rm v}\Omega\hat{z}\times \left(\rho_0{\rm v}- \rho_{\rm v}{\rm u}\right).\label{Eq:Magnus}
\end{equation}
The term in brackets is written in such a way that the force stays constant in order that fluctuations in the zero energy states are bounded. Replacing ${\rm v}$ with equation (\ref{Eq:Flux}) and using $\rho_{\rm v} F = \dot{\rm P} = -{\rm i}\omega m_{\rm v}\rho_{\rm v} {\rm v}$ up to leading order in gradients expansion, we get
\begin{equation}
  B_{\rm v}(\omega) = \frac{B_0\big[1 + (\xi/2)(k\ell)^2\big]}{(\omega/\Omega)^2 - \big[1 +(\xi/2)(k\ell)^2\big]^2}.\label{Eq:LastOne}
\end{equation}

\section{Results, Discussion and Conclusion\label{Sec:Section3}}
We now discuss the behavior of anomalous viscosity. 

Here, we observed how small temperature gradients affect viscosity of vortex fluid quantized on Hall states having filling fraction within $0<\nu_G<2$.

\begin{figure}[th!]
\centering
  \includegraphics[width=8.5cm,height=5.0cm]{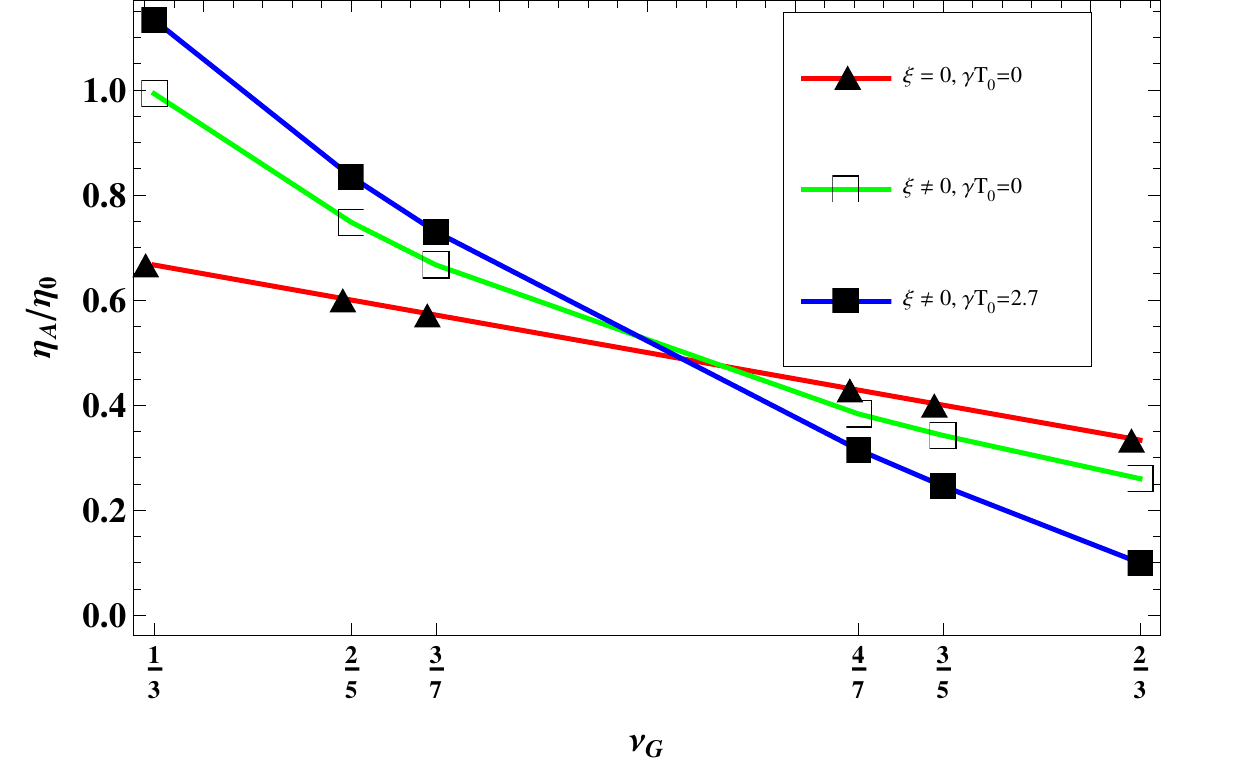}
  \includegraphics[width=8.55cm,height=5.50cm]{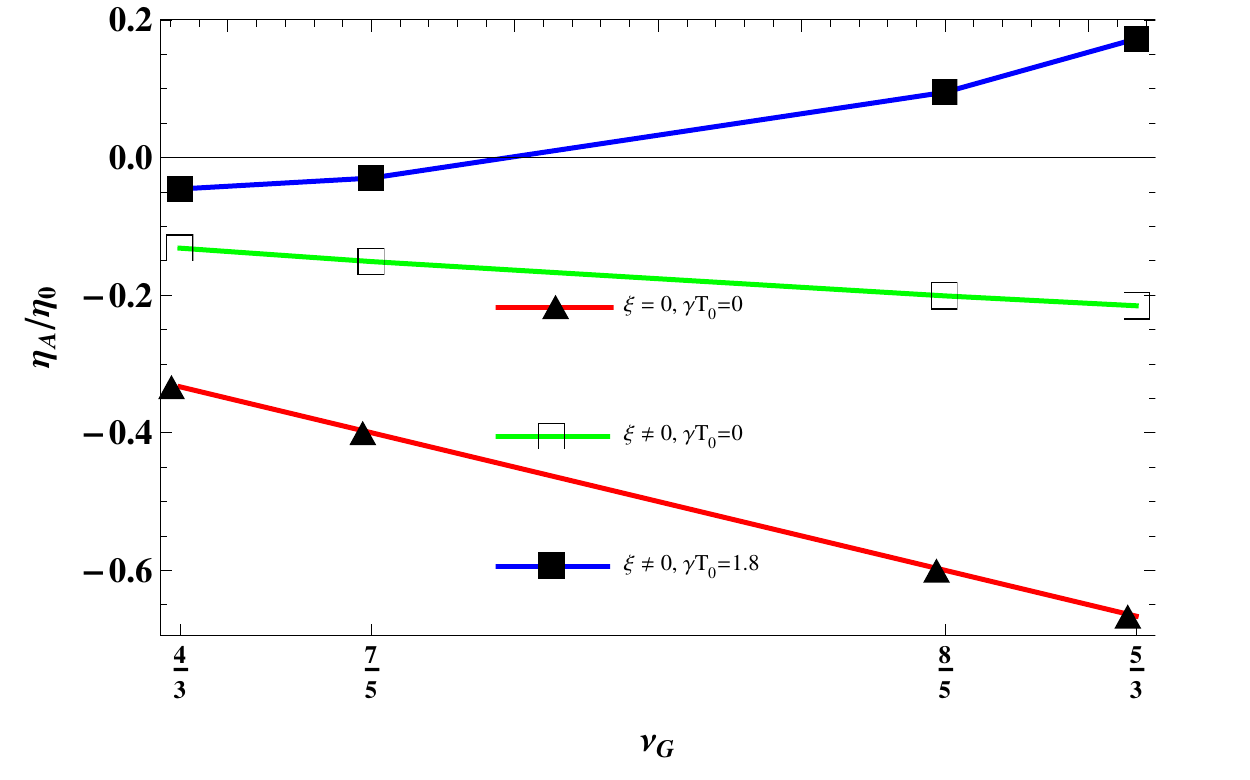}
\caption{Temperature gradient enhanced observations of weak fractions. $T_0 = 4.0\,K$, $\gamma = 0.9$ $\omega/\Omega = 0.1$ and $k\ell^2 = 0.2$. (top) $0 < \nu_G < 1$ and (Buttom) $ 1 < \nu_G < 2$.}\label{Fig:One}
\end{figure} 

In Figure 1 (a), the deviation of the $\eta_A$-curves from the mean viscosity, $\eta_A=0.55\eta_{0,A}$ measures the observability (strength) of a Hall state within $0<\nu_G<1$. It is clear that both $\nu_G=1/3$ and $\nu_G=2/3$ give the highest deviations which collaborate recent experimental results. See Refs~\cite{Dean2011, Amet2014} and references there in. Figure 1 (b) is similarly obtained for fractions within $1< \nu_G<2$, except that the mean viscosity is $\eta_A=- 0.55\eta_{0,A}$. The negative viscosity may suggest why these fractions are weakly observed in experiments.

\begin{figure*}[h!]
\centering
 \includegraphics[width=7.6cm,height=6.0cm]{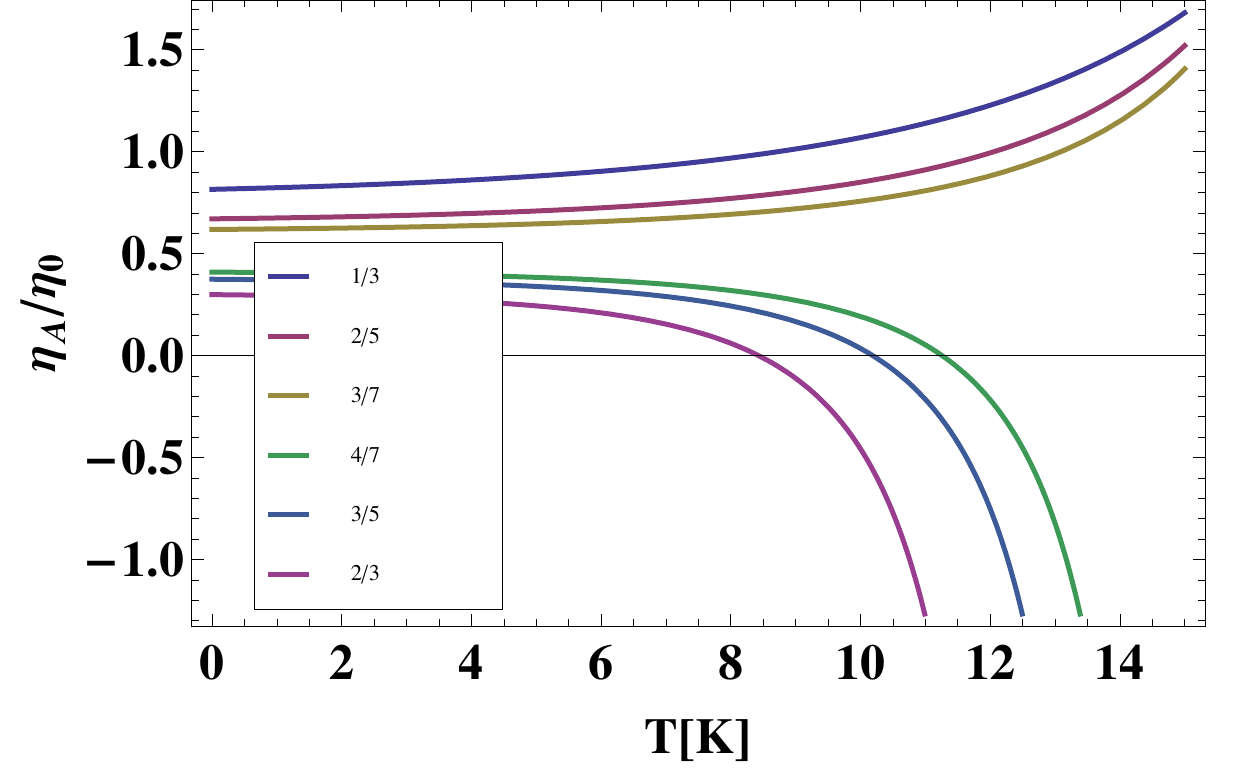}
 \includegraphics[width=7.6cm,height=6.0cm]{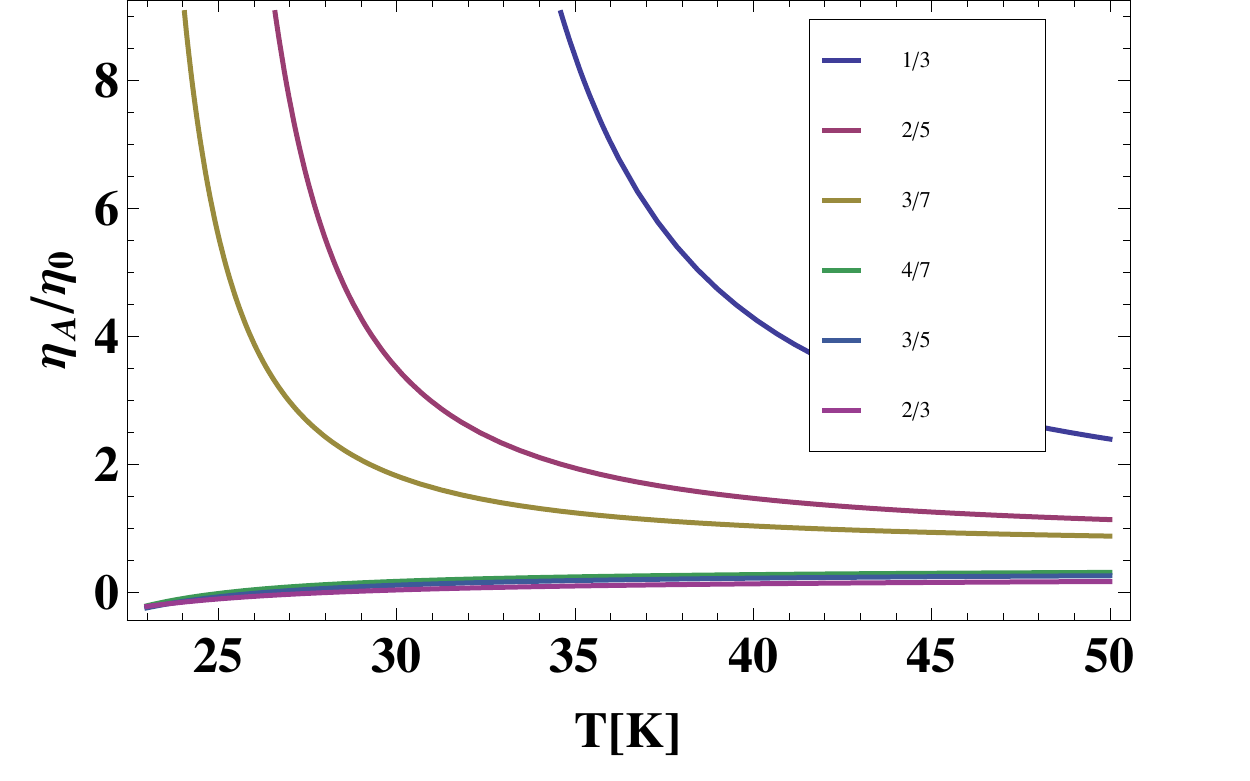}
 \includegraphics[width=7.6cm,height=6.0cm]{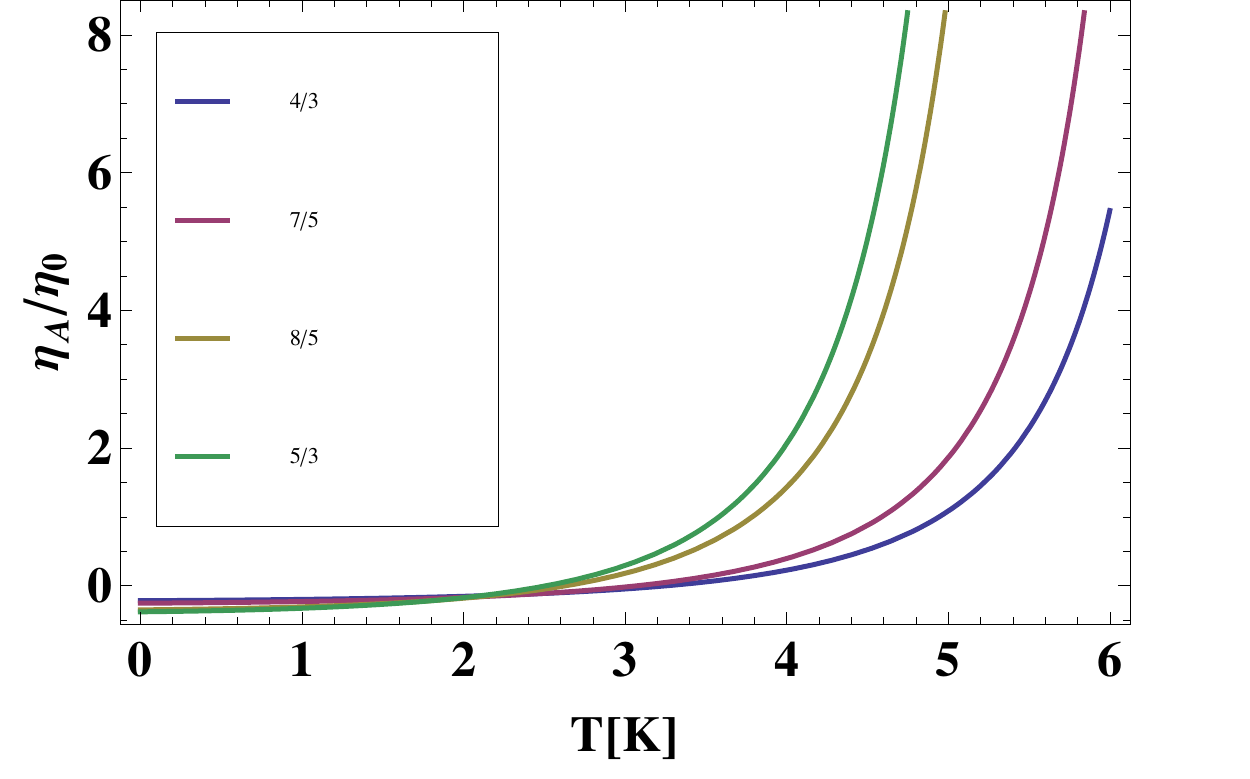}
 \includegraphics[width=7.6cm,height=6.0cm]{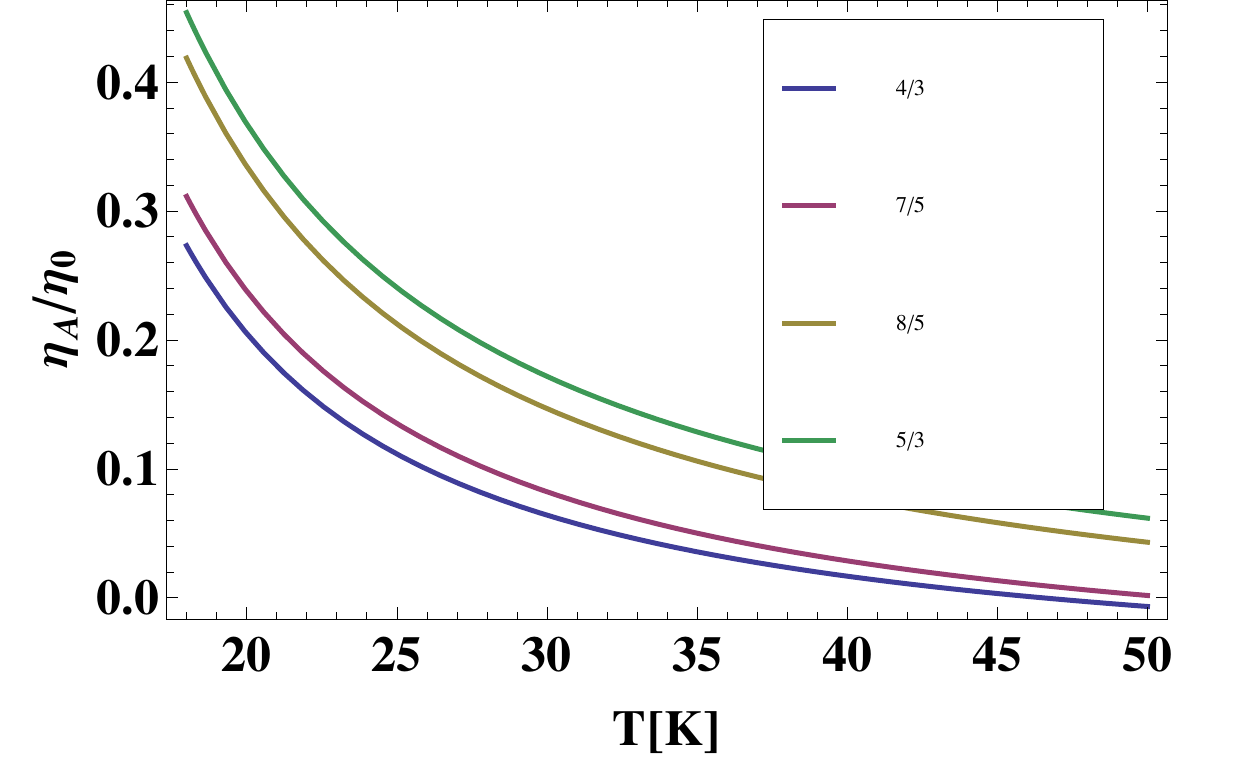}
\caption{Normalized anomalous viscosity plots versus temperature at different fractions showing infinities corresponding to critical temperatures. $\gamma = 0.9$ $\omega/\Omega = 0.1$ and $k\ell = 0.2$.  (top) $0 < \nu_G < 1$ and (Buttom) $ 1 < \nu_G < 2$.}\label{Fig:Two}
\end{figure*} 

In Figure 2 (a) and (b), we observe the anomalous viscosity over varying temperature, $T$. At some critical values, $T_C$, the viscosity grows to positive and negative infinities for $\nu_G>3/7$ and $\nu_G > 4/7$, respectively. Figure 2 (c) and (d) was similarly obtained for fractions within $1 < \nu_G < 2$.  Above and below $T_C$, $\eta_A$ is robust (flat). This particular character places graphene as an important industrial material where temperature and viscosity are of utmost importance.

\begin{figure*}[h!]
\centering
 \includegraphics[width=8.0cm,height=7.cm]{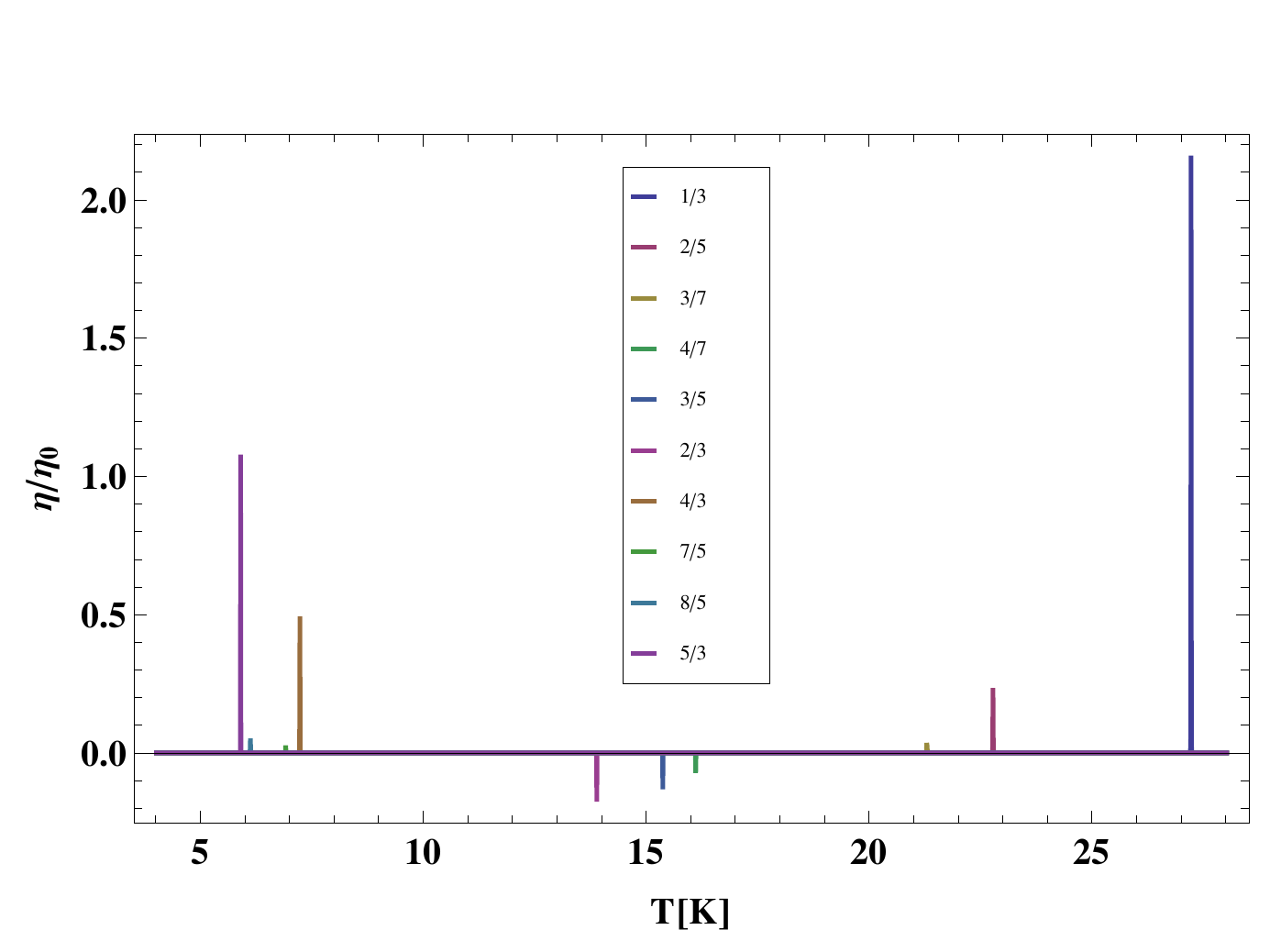}
 \includegraphics[width=8.8cm,height=6.5cm]{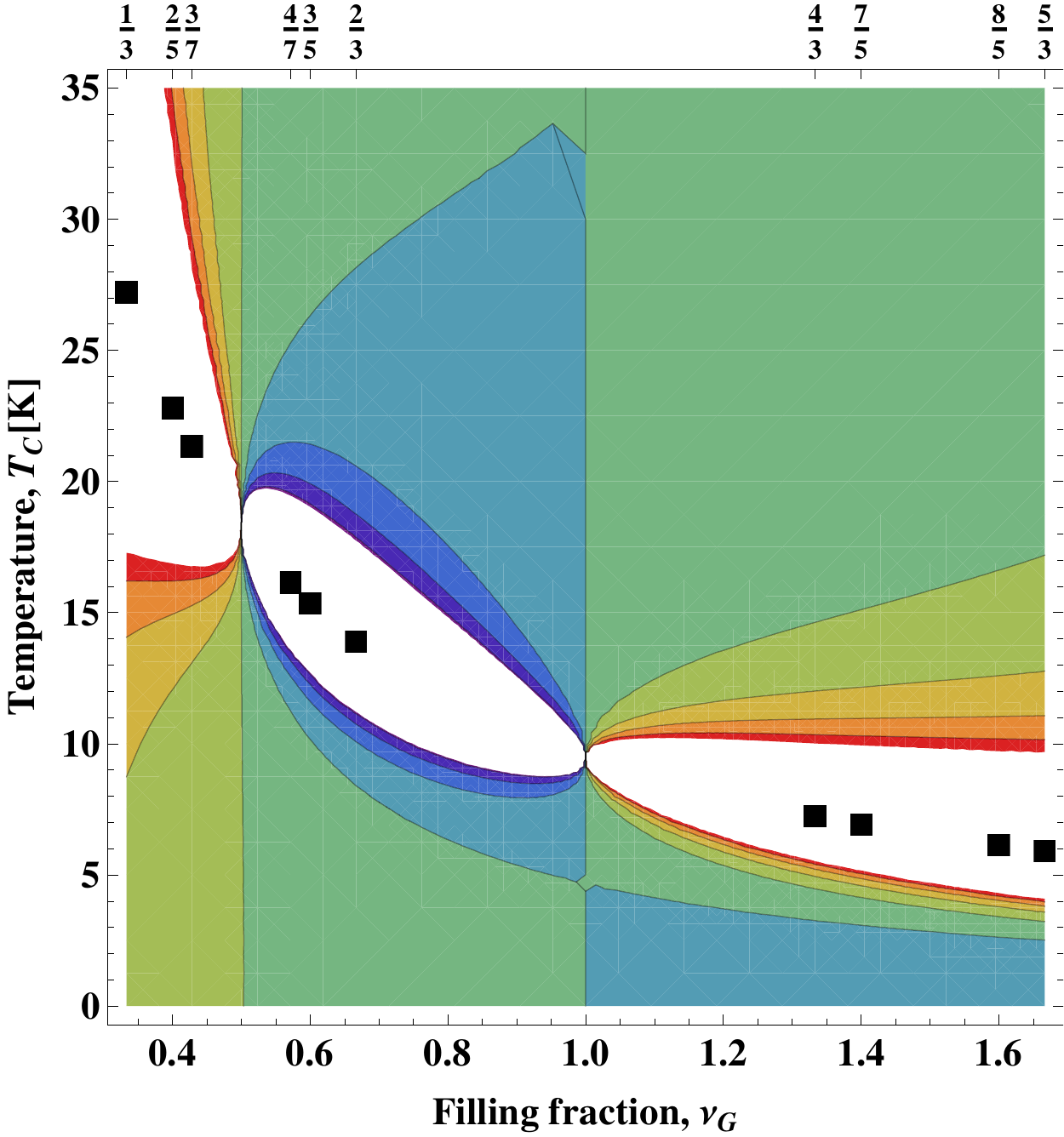}
 \caption{(Left) Normalized viscosity-temperature plot. Poles correspond to critical temperatures, $T_C$. (Right) $\eta_A$-T$_C$-$\nu_G$-plot. Full squares indicate critical temperatures. $\gamma = 0.9$ $\omega/\Omega = 0.1$ and $k\ell = 0.2$.} \label{Fig:Three}
\end{figure*}

In Figure \ref{Fig:Three2} we plotted $\eta$ in the neighborhood of infinites initially excluded from Figure \ref{Fig:Two}. We extracted the critical temperatures $T_C$ giving rise to $\eta_A$ infinities  and study $\eta$(T$_C$, $\nu_G$) behavior as shown in Figure \ref{Fig:Three}. The following characterized the appearance of observed empty spaces in the plot. The wider empty spaces (containing black squares) mean many fractions still remain hidden which experiments have not yet measured. With improved precision and sensitivity, experiments should be able to observe these fractions.

The intuitive meaning of the infinities in $\eta_A$ can be clarified further. The infinities correspond to, up to a factor of some function, $f(\nu_G)$ of the filling factor, the activation energy gap $\Delta_{\nu_G}$. To estimate the energy gap, we write $\Delta_{\nu_G} = f(\nu_G)T_C$ and expand $f(\nu_G)$ in the neighborhood of one of the transition zones in Figure \ref{Fig:Three}; $\nu_G\sim 1/2$ and $\nu_G\sim 1$. After applying some approximations, we obtained $f(\nu_G) = \lambda_{\nu_G}\nu_G(1-\nu_G)$. Where $\lambda_{\nu_G}$ is the degeneracy of $\nu_G$-fraction. For example, for $\nu_G=1/3$, $\lambda_{1/3} = 1$ and $\Delta_{1/3} = 18.14\,K$. Similarly, one has $\Delta_{2/5} = 5.47\,K$, $\Delta_{3/7} = 5.22\,K$, $\Delta_{4/7} = 3.95\,K$, $\Delta_{3/5} = 3.68\,K$, $\Delta_{2/3} = 9.25\,K$. These values agreed favorably with recent experiments~\cite{Dean2011, Amet2014}.

In conclusion, we have computed dissipationless viscosity of quantum vortex Hall states in graphene within hydrodynamics using quantum many-vortex picture of Euler hydrodynamics. The hydrodynamics formalism allowed a great deal of simplifications as the microscopic theory is completely unnecessary and only few variables, $\rho$ and ${\rm v}$ are employed.

An important aspect of the dissipationless viscosity are the anomalous $\xi$ and Hall $\gamma$ terms. Their effects come in two folds. (i) They reconstruct vortex density by introducing non-linearities in the fluid dynamics making flows at different FQHS distinguishable. (ii) $\xi$ captures temperature variations due to strain effects through $\gamma$ as soon as vortices are formed. The combined effect make viscosity robust over wider temperature range and can even enhanced observation of weaker FQH states. Controlling these parameters can enhance observability of weaker fraction and expose weaker ones. Knowing these parameters may be critical in strained-engineered devices. Since device sensitivity to stress-preserving deformation depends on the factors. 

$\eta_A$ contains infinities within some specific temperature ranges. These temperatures are associated with energy gaps. We estimated this energy gaps within $0 < \nu < 2$ FQH states by analysing the temperature dependence of $\eta_A$. $\xi$ and Hall $\gamma$ terms shift $\Delta_{\nu}$ to higher values. We found $\Delta_{1/3} = 18.14\,K$ that has been difficult to measure or mostly underestimated in experiments. Our estimated gaps also compares favorable with literature. See Refs.~\cite{Amet2014, Dean2011} and references there in.

Finally, our results could be applied to strained-engineered devices to control viscosity. The studies can also guide future experiments towards observing new fractions. In particular, the temperature windows (white spaces in Figure \ref{Fig:Three}) obtained may be probed, though away from the transition zones, for new fractions by controlling the Hall expansion coefficient parameter. Moreover, our work may resolve conflicts of different reported energy gaps. Specifically, the $\Delta_{1/3}$ gap.

\begin{acknowledgments} 
Rabiu will like to thank International Center for Theoretical Physics (ICTP)-Italy for hospitality and provision of travel grants to conduct part of the work at the center in Trieste.
\end{acknowledgments}


\end{document}